# Cluster richness–mass calibration with cosmic microwave background lensing

James E. Geach[1] & John A. Peacock[2]

[1]*Centre for Astrophysics Research, University of Hertfordshire, College Lane, Hatfield, AL10 9AB*

[2]*Institute for Astronomy, Royal Observatory, University of Edinburgh, Blackford Hill, Edinburgh, EH9 3HJ*

**Identifying galaxy clusters through overdensities of galaxies in photometric surveys is the oldest[1,2] and arguably the most economic and mass-sensitive detection method,[3,4] compared to X-ray[5–7] and Sunyaev-Zel'dovich Effect[8] surveys that detect the hot intracluster medium. However, a perrennial problem has been the mapping of optical 'richness' measurements on to total cluster mass.[3,9–12] Emitted at a conformal distance of 14 gigaparsecs, the cosmic microwave background acts as a backlight to all intervening mass in the Universe, and therefore has been gravitationally lensed.[13–15] Here we present a calibration of cluster optical richness at the 10 per cent level by measuring the average cosmic microwave background lensing convergence measured by *Planck* towards the positions of large numbers of optically-selected clusters, detecting the deflection of photons by haloes of total mass of the order $10^{14} M_\odot$. Although mainly aimed at the study of larger-scale structures, the *Planck* lensing reconstruction can yield nearly unbiased results for stacked clusters on arcminute scales. The lensing convergence only depends on the redshift integral of the fractional overdensity of matter, so this approach offers a clean measure of cluster mass over most of cosmic history, largely independent of baryon physics.**

Experiments such as the Atacama Cosmology Telescope (ACT),[16] South Pole Telescope (SPT)[17] and the *Planck*[18] satellite have now detected gravitational lensing of the cosmic microwave background (CMB) and produced large-area maps of lensing potential that detect features on 10-arcminute scales, effectively representing the projected mass density of the Universe back to the surface of last scattering. This offers a new method of understanding how visible matter traces the overall matter density field.[19,20] Clusters of galaxies are the rarest peaks in the matter density field and contain the most massive and oldest galaxies at any epoch; for this reason, they are simultaneously powerful cosmological probes[21] and unique environments for studying galaxy evolution. Naturally, clusters are expected to be strong deflectors of CMB photons, and therefore should be detectable, at least statistically, in maps of the CMB lensing field. This was first demonstrated with the $3.1\sigma$ detection of the lensing signal of 512 Sunyaev-Zel'dovich Effect (SZE) seleted clusters in SPT data,[22] where the lensing- and SZE-derived cluster masses were found to be in agreement. A $3.2\sigma$ statistical lensing signal was also detected in the ACT lensing map for 12,000 optically-selected galaxies, thought to reside on average in group-scale halos $M \approx 10^{13} M_\odot$,[23] demonstrating how the radial profile of the lensing convergence can be used to estimate the projected mass density around the target galaxies.[15,25] We now extend this to more massive systems over $1/3$ of the sky.

The convergence of the lensing field caused by a foreground mass can be expressed in the classical lensing framework in terms of the projected mass density as

$$\kappa(R) = \frac{\Sigma(R)}{\Sigma_{\rm crit}}, \quad (1)$$

where $R$ is the projected radius from the lens. $\Sigma_{\rm crit}$ is the critical mass density

$$\Sigma_{\rm crit} = \frac{c^2}{4\pi G} \frac{D_{\rm OS}}{D_{\rm OL} D_{\rm LS}}, \quad (2)$$

where $D_{OS}$, $D_{OL}$ and $D_{LS}$ are respectively the angular diameter distances between the observer and the source (in this case the surface of last scattering), between the observer and the lens and between the lens and the source. With a model for the mass density distribution one can predict the observed CMB lensing convergence field for any foreground structure (Methods).

The 'red sequence Matched-filter Probabilistic Percolation' (redMaPPer) algorithm[4] is one of the most successful 'red sequence' cluster detectors to date, and has been applied to the full Sloan Digital Sky Survey (SDSS) Data Release 8[26] photometric catalogue, detecting 26,111 clusters over approximately 10,500 square degrees. Central to redMaPPer is the richness parameter $\lambda$, which estimates the number of cluster member galaxies based on a sum over membership probabilities $p(\mathbf{x}|\lambda)$, where $\mathbf{x}$ are observable properties of a galaxy (such as colour). Since the number of cluster members should scale with total cluster mass, we expect to find a correlation between $\lambda$ and $\kappa$. The predicted convergence amplitude for individual clusters is undetectable in the *Planck* lensing map, but we consider the average convergence signal along the line of sight to a large sample of clusters. Although not without its problems, 'stacking' clusters has two advantages: (1) it allows one to detect a statistical signal at a level of order $\sqrt{N}$ below than the formal sensitivity of the lensing map, and (2) for independent targets, physically uncorrelated signal is averaged out, resulting in a cleaner measurement of the lensing profile.

We bin the total sample of 26,111 clusters into four richness classes determined by the quartile ranges of $\lambda$: [20,26), [26,33), [33,46), [46,302]. This approach allows us to sample the full range of richness while maintaining approximately equal Poisson errors for each bin, with approximately 6500 clusters per bin. There is a slight bias of increasing redshift with richness, with the median redshift spanning $\langle z \rangle = 0.32$–$0.45$ across the four bins (Table 1), but any evolutionary impact of the mass-richness relation over this range is expected to have a weak or negligible impact on our analysis. Our stacking approach is to reproject a 1 degree region of the *Planck* convergence map around the position of each cluster onto a tangential sky projection with a scale of 256 pixels per degree. For each richness bin we evaluate the mean ('stacked') local convergence map, ignoring masked values due to Galactic foregrounds and point sources, including cluster targets with *Planck*-detected SZE. The fraction of redMaPPer clusters masked in the *Planck* map is, respectively, 3.2, 3.2, 3.5 and 5.5 per cent of targets in each of the four richness bins. The resulting stacked average convergence maps are shown in Figure 1, indicating that the statistical convergence signal is significantly detected at peak signal-to-noise of 3.2, 4.2, 4.9 and 6.8$\sigma$ across the four richness bins, with the 1$\sigma$ noise level approximately $\sigma_\kappa \approx 0.01$ in each stack. This is a clear positive correlation, as expected for a progression in deflecting mass with richness.

The *Planck* convergence map is constructed using a quadratic estimator to separate the non-Gaussian lensing signal from the Gaussian CMB temperature and polarisation maps,[18,27,28] where lensing has the effect of correlating angular modes of the primordial power spectrum and introduces non-Gaussianity in the anisotropy.[15] An important systematic effect is therefore the presence of additional unmasked signal that also distorts the CMB, which could potentially bias the local estimate of lensing convergence.[14,22] Of most relevance for this work is the inverse Compton scattering of CMB photons with free electrons in the intracluster medium (i.e. the thermal SZE[8]). Although point sources are masked in the *Planck* reconstruction,[18] a statistical SZE signal from unresolved sources remains. This introduces a spectrally dependent bias at the per cent level for the reconstructed lensing power spectrum.[17,29] Since this signal is correlated with the positions of the very clusters that are being stacked, disentangling its contribution to the measured convergence amplitude is challenging. However, since the thermal SZE has a characteristic frequency dependence and the *Planck* lensing reconstruction combines several frequencies, including the SZE null at 217 GHz, as well as including polarisation information,[18] any low-level systematic effect caused by unmasked thermal SZE is small compared to the statistical uncertainty of the stacked convergence (we demonstrate this explicitly in the Methods). The kinetic SZE due to peculiar motions of clusters with respect to the CMB frame is frequency independent, uncorrelated with the primordial temperature anisotropy and largely Gaussian, so that any impact on the lensing reconstruction is negligible.[14]

To estimate the average cluster mass associated with each richness bin we construct model convergence maps assuming that the mass distributions of the clusters can be described by Navarro-Frenk-White profiles[30] (Methods). We adopt the conventional mass $M_{200}$, describing the mass within the radius $r_{200}$ that encloses a mass density of 200 times the critical density of the Universe, and assume a concentration parameter that scales with halo mass and redshift. With the redshift distribution of the sample known, we fit the radial average convergence profiles using $M_{200}$ as the single free parameter. In addition to the cluster halo, we allow for a further lensing signal due to large scale structure correlated with the clusters (the two-halo term), where the bias of the correlated overdensity can be described as a function of $M_{200}$ (Methods). The best fitting model profiles are shown in Figure 2 and Table 1 lists the peak convergence and best fit mass for each richness bin.

Figure 3 presents the $\lambda$–$M_{200}$ relation. As in other works, we model the relation as a power law $M_{200}(\lambda) = M_0(\lambda/\lambda_0)^\alpha$, with $\lambda_0 = 40$ fixed as the reference point. The best fitting parameters are $\log_{10}(M_0/M_\odot) = 14.39 \pm 0.05$ and $\alpha = 0.74 \pm 0.30$. This result is broadly in agreement with the mass calibration of redMaPPer clusters measured through stacked velocity dispersion,[11] stacked galaxy weak lensing[12] and the SZE.[9] The mass normalisation at $\lambda = 40$ is consistent within the errors for the four independent richness-mass calibrations; however, there is some tension in the slope $\alpha$, with the stacked CMB lensing result favouring a value consistent with that obtained via the SZE, predicting marginally higher halo masses compared to galaxy weak lensing and velocity dispersion at low $\lambda$. Nevertheless, the differences between the four methods are not significant with the current empirical constraints, and indicates the CMB weak lensing can provide a $\sim$10 per cent mass-richness calibration accuracy at $M_{200} \approx 2.5 \times 10^{14} M_\odot$. We can set a limit on the intrinsic dispersion of $M_{200}$ at fixed $\lambda$ by searching for excess variance in the distribution of $\kappa$ for the richest $\lambda$ bin, assuming a lognormal distribution[24] for $M_{200}(\lambda)$ (Methods). We measure no significant excess variance in $\kappa$ for the clusters contributing to the richest bin, with $\Delta\sigma_\kappa^2/\bar{\kappa}^2 = 3.4 \pm 4.1$, and thus estimate that the corresponding $3\sigma$ upper limit for $\Delta\sigma_\kappa^2$ would require an intrinsic dispersion in mass of $\sigma_{\log_{10}(M_{200})} < 0.9$ dex.

Reliable calibrations of optical richness to cluster mass will be increasingly important as deep and wide optical surveys increase in scale over the next decade and beyond. The Large Synoptic Survey Telescope is expected to detect hundreds of thousands of groups and clusters of galaxies out to $z \approx 1$, for example. While there are several methods to calibrate optical richness, including the weak lensing of background galaxies, CMB lensing provides a new, complementary approach,[25] offering at least three key advantages over other methods. The first is that, while not free of systematics (Methods), CMB lensing is arguably more robust than than galaxy weak lensing, suffering fewer systematics (e.g. shape measurements and photometric redshifts). The second is that lensing maps now exist for the majority of the extragalactic sky, with the surface of last scattering providing a backlight originating at a single redshift of $z = 1100$. The third is that the bulk of the CMB lensing signal is caused by structure at high redshifts ($z \approx 2$). Thus, cross-correlating maps of the CMB lensing field with clusters offers a self-consistent mass calibration technique for galaxy clusters observed across the bulk of cosmic history.[25]

## Contact

## Acknowledgements

The authors thanks G. Holder, A. Lewis, M. Madhavacheril, P. Marshall and E. Rozo for helpful discussions. J.E.G. is supported by a Royal Society University Research Fellowship.

## Author Contributions

Both authors contributed equally to the analysis and writing of the manuscript.

# Figures and Tables

Table 1: Peak average convergence and best fit halo mass for the four richness bins defined in this study. The four bins represent the quartile ranges of $\lambda$, with $\langle\lambda\rangle$ giving the median richness and $\langle z\rangle$ the median redshift of each bin. The errors on $\langle\lambda\rangle$ and $\langle z\rangle$ indicate the 5–95 percentile range for those distributions. We also give the fit quality in terms of $\chi^2$/dof.

| $\lambda$ | $\langle\lambda\rangle$ | $\langle z\rangle$ | $\bar{\kappa}/10^{-3}$ | $\log_{10}(M_{200}/M_\odot)$ | $\chi^2$/dof |
|---|---|---|---|---|---|
| 20–26 | $23^{+3}_{-3}$ | $0.32^{+0.08}_{-0.17}$ | $35\pm11$ | $14.16^{+0.12}_{-0.15}$ | 0.40/6 |
| 26–33 | $29^{+4}_{-3}$ | $0.37^{+0.08}_{-0.20}$ | $42\pm10$ | $14.27^{+0.09}_{-0.11}$ | 1.11/6 |
| 33–46 | $38^{+6}_{-5}$ | $0.41^{+0.09}_{-0.23}$ | $49\pm10$ | $14.31^{+0.09}_{-0.11}$ | 2.54/6 |
| 46–302 | $58^{+46}_{-11}$ | $0.45^{+0.11}_{-0.25}$ | $61\pm9$ | $14.50^{+0.06}_{-0.07}$ | 6.47/6 |

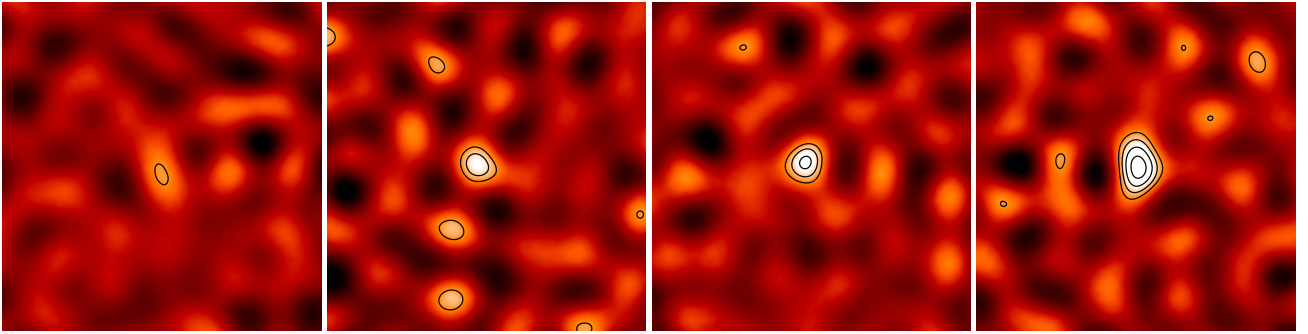

Figure 1: The average convergence of the CMB lensing field in the direction of clusters in bins of increasing optical richness ($\lambda$). Each image spans 1 degree, and the resolution of the convergence map is approximately 5 arcminutes. Contours show the significance of the signal starting at $3\sigma$ and increasing in steps of $1\sigma$, which we measure as the standard deviation of the equivalent stacks performed in noise maps. We bin the clusters by richness quartile, so the noise is roughly equivalent in each map, with $\sigma_\kappa \approx 0.01$.

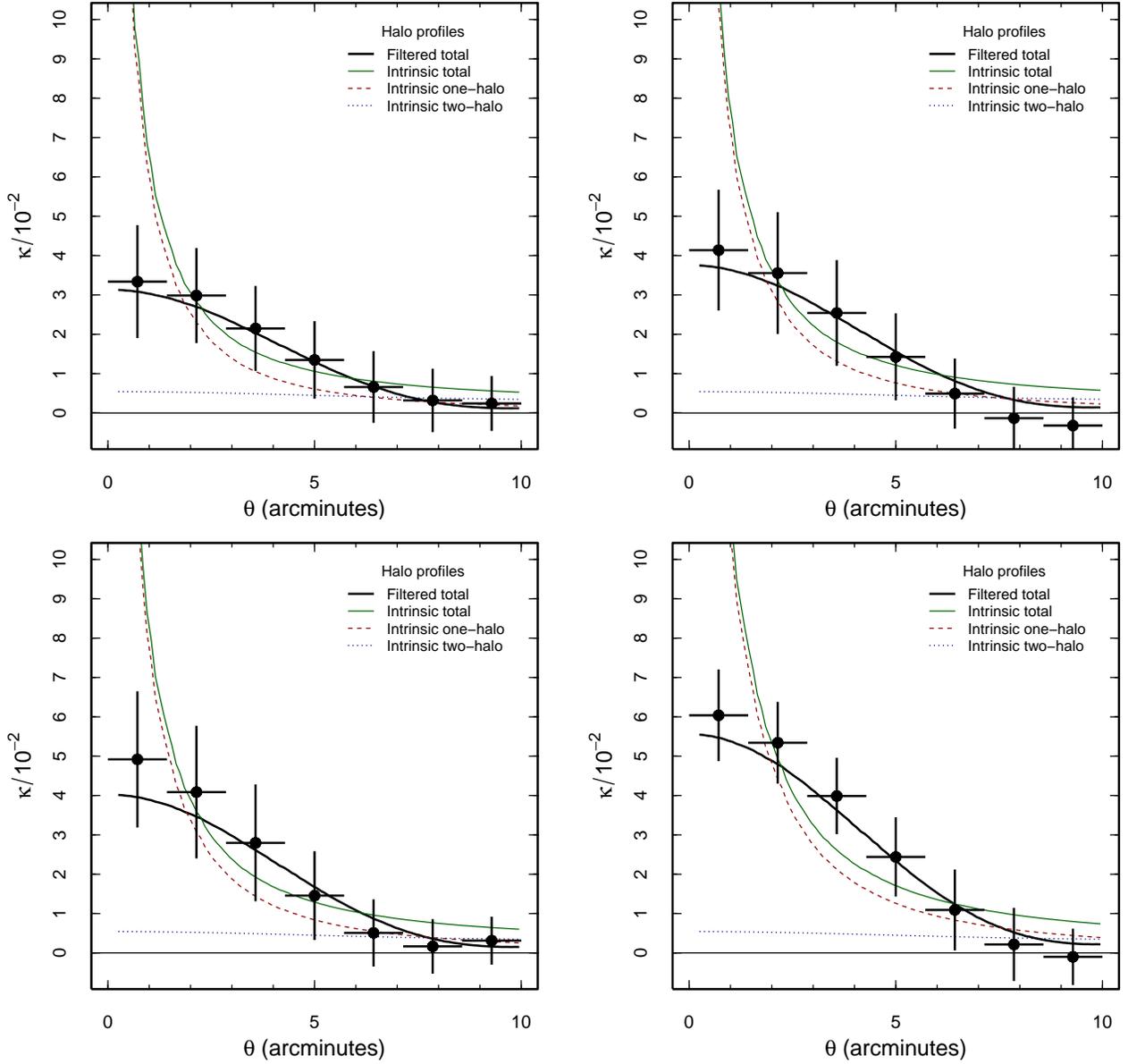

Figure 2: Radial profiles of the average convergence towards clusters in bins of optical richness ($\lambda$). The points show the azimuthally averaged values within 10 arcminutes of the centres of the maps shown in Figure 1 with uncertainties derived from the variance of equivalent stacks in noise simulations. Lines show the best fitting model consisting of a NFW density profile plus two-halo term to account for correlated matter, and the intrinsic convergence profile (split into one and two-halo contribution) expected in the absence of the low pass filtering that has been applied to the lensing map and model.

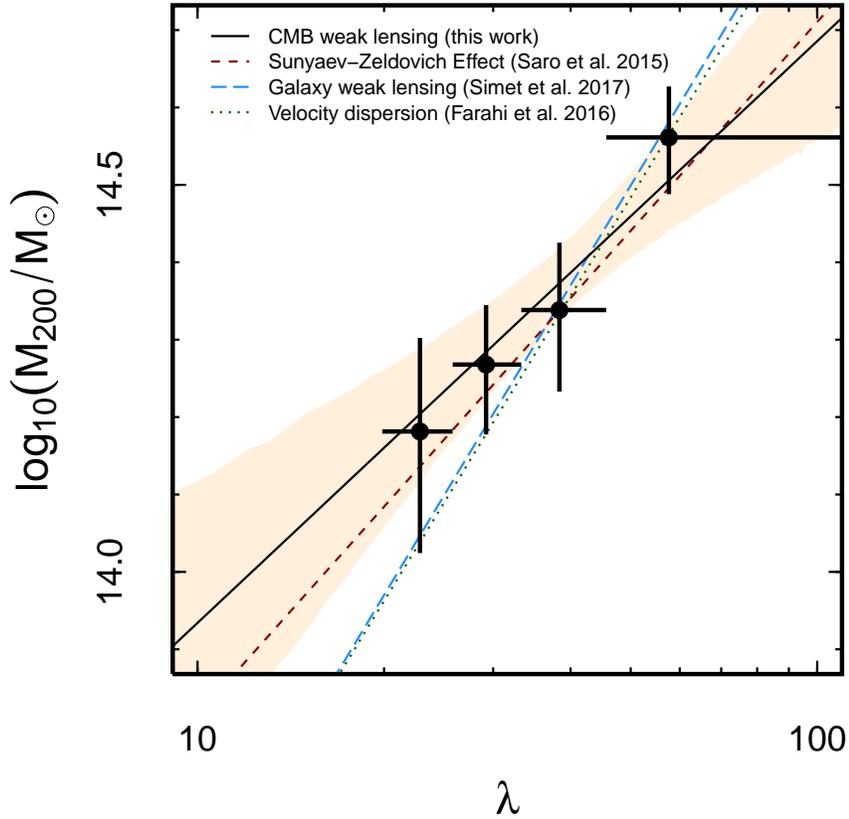

Figure 3: The mass-richness relation for stacked CMB lensing. Vertical error bars show the $1\sigma$ confidence range for the best fit mass, horizontal error bars show the richness range. All literature masses have been converted to $M_{200}$ in terms of the critical density of the Universe and matched in background cosmology. The SZE scaling includes an evolutionary term which we have applied for redshift evolution to $z = 0.4$. The shaded region indicates the range of acceptable fits to the relation $M_{200}(\lambda) = M_0(\lambda/\lambda_0)^\alpha$ within $1\sigma$ of the best fitting parameters.

# Methods

## Lensing framework

The projected mass density for a three dimensional mass density distribution $\rho(r)$ is given by

$$\Sigma(R) = 2 \int_R^\infty \frac{r\rho(r)}{\sqrt{(r^2-R^2)}} dr \qquad (3)$$

We consider a Navarro-Frenk-White (NFW) density profile[30] for clusters

$$\rho(x) = \frac{\rho_s}{x(1+x)^2} \qquad (4)$$

where the radial dependence is given in units of a scale radius, $x = r/r_s$. We consider a standard scale radius $r_s = r_{200}/c$, where $r_{200}$ is the radius at which the enclosed mass density is equal to 200 times the critical density of the Universe at a given epoch, $\rho_{\rm crit}(z)$, and $c$ is the halo concentration parameter. We adopt the mass and redshift dependent scaling for the concentration[31]

$$c(M_{200}, z) = 5.71(1+z)^{-0.47} \left(\frac{M_{200}}{2 \times 10^{12} h^{-1} M_\odot}\right)^{-0.084} \qquad (5)$$

where $M_{200}$ is the mass within $r_{200}$

$$M_{200} = \frac{800\pi}{3} \rho_{\rm crit}(z) r_{200}^3. \qquad (6)$$

The relevant density to be used in equation (4) is then[30,32]

$$\rho_s = \frac{200\rho_{\rm crit}(z)}{3} \frac{c^3}{\ln(1+c) - c(1+c)^{-1}}. \qquad (7)$$

We also expect a contribution to the lensing signal from large scale structure correlated with the clusters, which is normally modelled as a 'two-halo' term[33–35]

$$\kappa_2(\theta) = \int \frac{\ell d\ell}{2\pi} J_0(\ell\theta) \frac{\bar{\rho}(z) b_h(M) P_m(k,z)}{(1+z)^3 \Sigma_{\rm crit} D_A^2(z)} \qquad (8)$$

where $J_0$ is the zeroth order Bessel function, $D_A(z)$ is the angular diameter distance, $P_m$ is the linear matter power spectrum (with $k = \ell/(1+z)D_A$), $\bar{\rho}(z)$ is the average density of the Universe at $z$ and $b_h(M)$ is the bias for a halo of mass $M$. We assume $b_h = f(\nu)$, where $\nu$ is the ratio of the critical threshold for spherical collapse to the root mean squared density fluctuation for a halo of mass $M$, and $f$ is a functional fit derived from $N$-body simulations[36] to evaluate $b_h$, which assumes the same convention for halo mass, $M_{200}$. Figure 2 shows the relative contribution of the two-halo term on the convergence profile, illustrating the non-negligible signal relative to the one-halo term at scales of ∼5 arcminutes.

Having established this framework for the expected CMB lensing convergence for a given cluster we have a means of fitting observations. We are applying a stacking analysis that averages over a large number, $N$, of clusters belonging to some redshift distribution $dn/dz$, such that the predicted convergence profile for an average halo mass $M_{200}$ is

$$\langle \kappa(M_{200}, \theta) \rangle = \sum_{i=0}^N [\kappa_1(M_{200}, \theta, z_i) + \kappa_2(M_{200}, \theta, z_i)] \left.\frac{dn}{dz}\right|_i \qquad (9)$$

where $\kappa_1$ is the one halo term defined by equations (1–4) and the integral of $dn/dz$ is normalised to unity. Throughout this work we have assumed a $\Lambda$CDM cosmology with parameters set by the *Planck* 2015 constraints.[37]

**Model fitting**

We use the 2015 version of the *Planck* lensing map.[18] The lensing map is constructed using a minimum variance quadratic esimator that aims to determine the lensing signal assuming its effect on the primordial temperature (*T*) anisotropy along a line of sight $\hat{\mathbf{n}}$ can be described as

$$T^{\text{observed}}(\hat{\mathbf{n}}) = T^{\text{primoridal}}\left(\hat{\mathbf{n}} + \nabla \phi(\hat{\mathbf{n}})\right) \qquad (10)$$

where $\phi$ is the lensing potential.[18,28] The 2015 *Planck* lensing map combines five estimators involving correlations of the CMB *T* and polarisation (*E,B*): *TT*, *TE*, *EE*, *EB* and *TB*, covering approximately 70% of the sky. To create stacked convergence maps we reproject the *Planck* Hierarchical Equal Area isoLatitude Pixelization (HEALpix) map to a tangential flat sky projection around the position of each cluster with a scale of 256 pixels per degree. To determine the null signal and measure the significance of our stacked detections, we stack at identical cluster positions in 100 independent random noise realisations of the convergence map.

We evaluate the model described by equation (7) on an equivalent pixel grid as the stacked maps to create theoretical stacked 2D convergence profiles. Two filtering steps are required to compare to the data: one is a convolution that accounts for miscentering of the redMaPPer clusters, since the galaxy identified as the central member can be misclassified, resulting in an offset from the centre of the potential well.[4] This miscentering has a smoothing effect on the stacked convergence profile ($\kappa \to \kappa^\star$), which in Fourier space can be expressed as[35]

$$\kappa_\ell^\star = \left[f_{\text{cen}} + (1 - f_{\text{cen}})\exp\left(-0.5\sigma_c^2 \ell^2\right)\right]\kappa_\ell \qquad (11)$$

where we adopt $f_{\text{cen}} = 0.8$ and $D_A(z)\sigma_c = 0.42 h^{-1}$ Mpc following other works.[4,35] The second filtering step is simply a low pass filter that removes power on scales of $\ell > 2048$, matching the filtering applied to the reconstructed convergence map.

To estimate the average mass, $M_{200}$, of each richness class, we measure the azimuthally averaged radial profile of the stacked maps ($\kappa_{\text{obs}}$) and models ($\kappa_{\text{model}}$) within a radius of 5 arcminutes. We fit the profiles with a Markov Chain Monte Carlo (MCMC) approach using *emcee*,[38] which implements an Affine Invariant Ensemble sampler. The fit optimises the log likelihood $\ln(\mathcal{L}) = -0.5\chi^2$ where

$$\chi^2 = (\kappa_{\text{obs}} - \kappa_{\text{model}})^{\text{T}} \mathbf{C}^{-1} (\kappa_{\text{obs}} - \kappa_{\text{model}}) \qquad (12)$$

where $N = 100$ and $i, j$ are bin indices and $\mathbf{C}$ is the stacked covariance matrix

$$\mathbf{C}_{ij} = \frac{1}{N-1} \sum_{k=0}^{N} (\kappa_i^k - \bar{\kappa}_i)(\kappa_j^k - \bar{\kappa}_j) \qquad (13)$$

where $N = 100$ and $i, j$ are the bin indices. We assume a flat prior for $M_{200}$ with $10^{13} < (M_{200}/M_\odot) < 10^{15}$ and derive 1$\sigma$ confidence bounds from the 16th and 84th percentiles of accepted samples for 100 walkers over 750 iterations (1000 in total, with the first 250 rejected as burn-in).

To estimate the intrinsic dispersion in $M_{200}$ in a given richness bin, we consider the excess variance

$$\Delta \sigma_\kappa^2 = \sigma_{\kappa,\text{measured}}^2 - \langle \sigma_{\kappa,\text{noise-only}}^2 \rangle \qquad (14)$$

where $\sigma_{\kappa,\text{measured}}^2$ is the variance in the distribution of peak $\kappa$ measured at the position of clusters contributing to a richness bin, and $\langle \sigma_{\kappa,\text{noise-only}}^2 \rangle$ is the equivalent measurement averaged over 100 noise-only realisations of the map as described above. The fractional uncertainty on $\sigma^2_{\kappa,\text{measured}}$ is given by $(2/(N-1))^{1/2}$, where $N$ is the number of unmasked clusters in the bin. We find no significant excess variance in the convergence values measured at the position of clusters in the richest bin, with $\Delta\sigma^2_\kappa/\bar\kappa^2 = 3.4 \pm 4.1$. We can set limits on the intrinsic dispersion in $M_{200}$ for the clusters contributing to the richest bin by assuming an intrinsic halo mass distribution at approximately fixed richness and finding the scale of the dispersion at which we would have detected a significant excess variance. For the same clusters contributing to the richest bin we generate a set of model $\kappa$ values using the same model described above and drawing individual $M_{200}$ randomly from a lognormal distribution. By increasing the scale of the mass distribution we can determine at what point we would have measured a significant ($3\sigma$) excess variance in the convergence distribution; we determine this limit to be $\sigma_{\log_{10}(M_{200})} < 0.9$ dex.

Our stacking analysis evaluates the average lensing mass for a sample of clusters with a range of richness. We have measured the mean mass for objects in relatively narrow ranges of richness. The actual bin width is irrelevant, because we report the median richness in the bin rather than the bin centre. However, it is nevertheless true that, for a set of richness-selected clusters, there is the scope for a Malmquist bias in the richness values. The standard value for this bias is

$$\Delta \ln \lambda = \sigma^2 \left( \frac{d \ln N}{d \ln \lambda} \right) \tag{15}$$

where $N$ is the integral counts and $\sigma$ is the measuring-error dispersion in $\ln \lambda$. For example, for a typical count slope of 2, and assume $\sigma = 0.2$. Then the values of $\lambda$ in a $\lambda$-selected sample would then be about 10% higher than their average true value. This is roughly the precision with which we are able to measure the mean mass at a given richness, and moreover richness varies roughly linearly with mass. Therefore, we can conclude that one might need to apply a different mass-richness calibration to richness-selected clusters (the current case) and to clusters selected in a way that is independent of richness, depending on whether or not the single-object scatter in the mass-richness relation is greater than 20%. From the current data, we cannot rule out a scatter this large or greater, so the possibility of Malmquist bias certainly exists, and the current calibration should thus be taken to apply only to richness-selected samples.

We fit the relation

$$\log_{10}(M_{200}/M_\odot) = \log_{10}(M_0/M_\odot) + \alpha \log_{10}(\lambda/\lambda_0) \tag{16}$$

with $\log_{10}(M_0/M_\odot)$ and $\alpha$ as free parameters using the same MCMC code as above, with $1\sigma$ confidence bounds again defined by the 16th and 84th percentiles of accepted samples for 100 walkers over 750 iterations after a burn-in of 250 iterations. The best fit parameters are $\log_{10}(M_0/M_\odot) = 14.39 \pm 0.05$ and $\alpha = 0.74 \pm 0.30$. The best fit $\lambda$–$M_{200}$ calibration and $1\sigma$ range of acceptable fits are shown in Figure 3.

**Systematics**

There are several sources of systematic error that could affect our measurement. First we consider the effect of masking clusters with significant SZE detections that were excluded from the lensing reconstruction; this will generally remove the most massive clusters in each richness bin. We use the redMaPPer-matched *Planck* SZE catalogue to determine what clusters have been excluded from the analysis in each richness bin, finding 2, 0, 18 and 280 matches. If we consider $M_{\text{lensing}} \approx M_{\text{SZE}}$, the inclusion of these masked clusters would change the average cluster mass by 2.7% in the richest bin, and ≲0.1% in the three poorer bins. We conclude that cluster masking has a negligible impact on our analysis.

Another concern is the impact of thermal SZE (tSZE) on the lensing reconstruction that could potentially bias estimates of $\kappa$ towards the direction of clusters. We test this by first stacking the *Planck* SMICA CMB temperature map at the same positions as our target clusters. We indeed find a richness dependent temperature decrement which peaks at $T_{\rm CMB} \approx 2\,\mu{\rm K}$ in the richest bin. Using this as a model profile for the typical level of tSZE contamination in the temperature map used to derive the lensing estimate, we perform a simple test: first we create an ensemble of 6500 2 degree CMB temperature flat-sky maps via random Gaussian realisations using the CMB power spectrum, then distort these using the best fitting convergence profile defined above for each richness bin. We do this for a pure CMB and one including the model empirical tSZE profile derived from the SMICA stack. We then apply a lensing estimator to each individual map and then average these to create two model $\kappa$ stacks with and without tSZE contamination (which is added after the distortion has been applied). In practice we apply the Hu (2001) estimator[39] based on $\nabla \mathcal{W} \vec{\mathcal{G}}$, where $\mathcal{W}$ and $\vec{\mathcal{G}}$ are suitably filtered versions of the temperature and temperature gradient (where we follow the approach of Hu, DeDeo & Vale 2007).[40] The simulated lensed maps have noise added to them in the form of white noise set at the approximate level of *Planck*, with the noise power spectrum equal to that of the intrinsic CMB at $\ell = 3000$. The reconstruction is low passed filtered like the real data, rejecting modes $\ell > 2048$. When we stack the 6500 model maps we recover the input model convergence amplitude and profile across all richness bins (Figure S1). The difference between the 'pure' and 'tSZE contamination' stacked convergence map does show a residual signal, indicating that the peak amplitude of the convergence estimate when tSZE is included is underestimated at the 0.5–2% level across our richness bins, increasing with halo mass. This bias is much smaller than our statistical error, but since it is most pronounced on scales below $10'$, might be important to consider in high resolution lensing reconstructions even when the tSZE has been compensated for. Figure S1 shows the radial profile of $(\kappa_{\rm tSZE} - \kappa_{\rm pure})/\kappa_{\rm pure}^0$ for each bin, where the denominator is the peak convergence measured at $\theta = 0$ in the pure reconstruction.

It has been argued that the reconstructed convergence estimated from the Hu non-linear filter[39] (designed for recovering the lensing signal on large scales) could underestimate the true convergence on scales below $\sim 2r_s$ at the positions of clusters[41]. The reason is that on the scales pertinent to cluster weak lensing, the CMB temperature anisotropy can be approximated as a temperature gradient, with the lensing signal preferentially recovered when the gradient is strong at the position of the cluster. In contrast, the lensing signal is poorly recovered when the temperature gradient across the cluster is low. If this is not accounted for (for example by statistically weighting), the convergence stack of a sample of clusters will be biased low because the convergence will not be recovered for a subset of the sample. To test for impact on our mass estimates, we model the bias in the lensing reconstruction as

$$\frac{\kappa_{\rm measured}}{\kappa_{\rm true}} = \left(\frac{r}{r_s}\right)^{0.23} \tag{17}$$

for scales $r < 2r_s$ and modify our projected halo model accordingly (corresponding to a factor of 2 underestimation of $\kappa$ at $0.1r_s$). Re-fitting the modified profiles yields no significant change in the mass estimate within the fitting uncertainties, and this is somewhat expected because the spatial scales at which the underestimate of the convergence is most pronounced lie at $\ell > 2048$, which are filtered out of the *Planck* reconstruction, and indeed this is confirmed by our simulations (Figure S1), indicating that any bias due to the lensing estimator is negligible in our analysis.

In this analysis we have chosen a redshift and mass dependent concentration scaling (equation 5) based on simulations, but how sensitive are the results to systematic changes in halo concentration? The $c(M,z)$ relation has also been measured empirically, with results from lensing studies of CLASH clusters[42] showing reasonable agreement with equation (5), but with $1\sigma$ empirical uncertainties of 10–25% over the redshift range of this study. We simulate the effect of a systematic shift in concentration of the same magnitude as the $1\sigma$ empirical uncertainties by generating $\kappa$ profiles using the framework described above and measuring the change in peak $\kappa$ compared to our nominal values, finding a variation of 6, 4, 3, and 2% in the best-fit mass for each richness bin. Again, this is small in comparison to our statistical 10% calibration uncertainty.

## Supplementary Figures

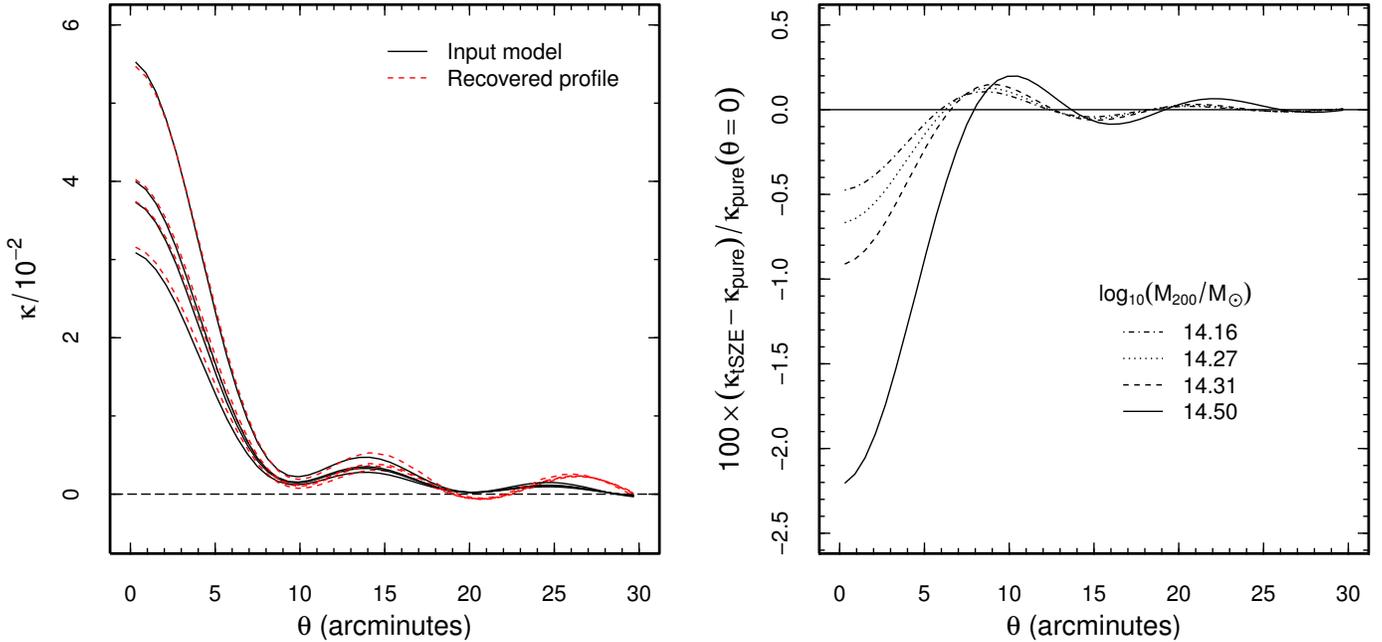

Figure S1: (left) Comparison of the convergence profiles of input models (solid lines) comprising the four best-fitting NFW profiles corresponding to each richness bin considered in this study, and estimates of the average convergence of 6500 CMB realisations each distorted with the input convergence model (red dashed lines). This demonstrates that the Hu et al. lensing estimator successfully recovers the stacked cluster lensing profile. Note that maps have been low passed filtered at $\ell > 2048$. (right) Comparison of the recovered convergence profiles for the lensing model with a pure CMB realisation and one where a residual tSZE signal commensurate with the same halo mass has been included. The presence of tSZE contamination causes the lensing estimator to underestimate the amplitude of the convergence at the 0.5–2% level, increasing to higher mass, and is negligible at large scales.